\documentclass[runningheads]{llncs}
\pdfoutput=1
\usepackage{graphicx}
\usepackage{ulem}
\usepackage{multirow}
\usepackage{color,graphicx}

\begin{document}
\title{Recognition and standardization of cardiac MRI orientation via multi-tasking learning and deep neural networks}
\titlerunning{Recognition and standardization of cardiac MRI orientation}
%
\author{Ke Zhang \and
Xiahai Zhuang\thanks{Xiahai Zhuang is corresponding author. This work was funded by the National Natural
Science Foundation of China (Grant No. 61971142), and Shanghai Municipal Science and Technology Major Project (Grant No. 2017SHZDZX01).}}
\authorrunning{K Zhang \& X Zhuang}
%
\institute{School of Data Science, Fudan University, Shanghai\\
\email{16307100128@fudan.edu.cn;zxh@fudan.edu.cn}}
\maketitle              
\begin{abstract}
In this paper, we study the problem of imaging orientation in cardiac MRI, and propose a framework to categorize the orientation for recognition and standardization via deep neural networks. The method uses a new multi-tasking strategy, where both the tasks of cardiac segmentation and orientation recognition are simultaneously achieved. For multiple sequences and modalities of MRI, we propose a transfer learning strategy, which adapts our proposed model from a single modality to multiple modalities. We embed the orientation recognition network in a Cardiac MRI Orientation Adjust Tool, i.e., CMRadjustNet. We implemented two versions of CMRadjustNet, including a user-interface (UI) software, and a command-line tool. The former version supports MRI image visualization, orientation prediction, adjustment, and storage operations; and the latter version enables the batch operations. The source code, neural network models and tools have been released and open via https://zmiclab.github.io/projects.html.
 
\keywords{Orientation recognition  \and Multi-task learning \and Cardiac MRI}
\end{abstract}
\section{Introduction}
Cardiac Magnetic Resonance (CMR) images could be stored in different image orientations when they are recorded in DICOM format and stored into the PACS systems. Recognizing and understanding this difference is crucial in deep neural network (DNN)-based image processing and computing, since current DNN systems generally only take the input and output of images as matrices or tensors, without considering the imaging orientation and real world coordinate.
This work is aimed to provide a study of the CMR image orientation, for reference to the human anatomy and standardized coordinate system of real world, and to develop an efficient method for recognition and standardization of the orientation.

Deep neural networks have been demonstrated to achieve state-of-art performance in many medical imaging tasks, such as image segmentation and lesion detection. For CMR images, standardization of all the images is a prerequisite for further computing tasks based on DNN-based methodologies, such as image segmentation ~\cite{ref_article9,ref_article10,ref_article13} and myocardial pathology analysis~\cite{ref_article6}. 

Nevertheless, recognizing the orientation of different modality CMR images and adjusting them into standard format could be as challenging as the further computing tasks. Different from other work that focuses on segmentation or classification individually~\cite{ref_article3} or combine image segmentation with quantification~\cite{ref_article12} this work proposes a DNN-based framework to solve the cardiac image segmentation and orientation recognition tasks simultaneously.

The original multi-tasking learning aims at exploiting commonalities and differences across tasks.  To extend this concept to deep learning, the multi-tasking framework trains the neural network to learn from different tasks and solve different medical image processing tasks at the same time. In recent years, multi-tasking methods in medical image processing have grown in popularity. Xue et al propose a  multitask learning network (FullLVNet)~\cite{ref_article8}, which modeled intra- and inter-task relatedness to enforce improvement of generalization. Vigneault et al presented the $\Omega$-Net (Omega-Net): a convolutional neural network (CNN) architecture for simultaneous localization, transformation into a canonical orientation, and semantic segmentation ~\cite{ref_article4}. The auxiliary task could improve the generalization performance by concurrently learning with the main task, which is the main merit if multi-task learning~\cite{ref_article1}. The previous work focuses on several segmentation tasks or deal with segmentation and classification tasks at the same time.  
To tackle the difference of CMR image orientation when they were presented for DNN-based image processing,  we propose a recognition task, and combine it with the traditional task of CMR image segmentation. 

Deep learning-based methods have been widely used in orientation recognition and prediction tasks. Wolterink et al proposed an algorithm that extracts coronary artery centerlines in cardiac CT angiography (CCTA) images using a convolutional neural network (CNN)~\cite{ref_article5}. Duan et al combine a multi-task deep learning approach with atlas propagation to develop a shape-refined bi-ventricular segmentation pipeline for short-axis CMR volumetric images~\cite{ref_article2}. Based on CMR orientation recognition, we further develop a framework for standardization and adjustment of the orientation.  

\begin{figure}[thb]
\includegraphics[width=\textwidth]{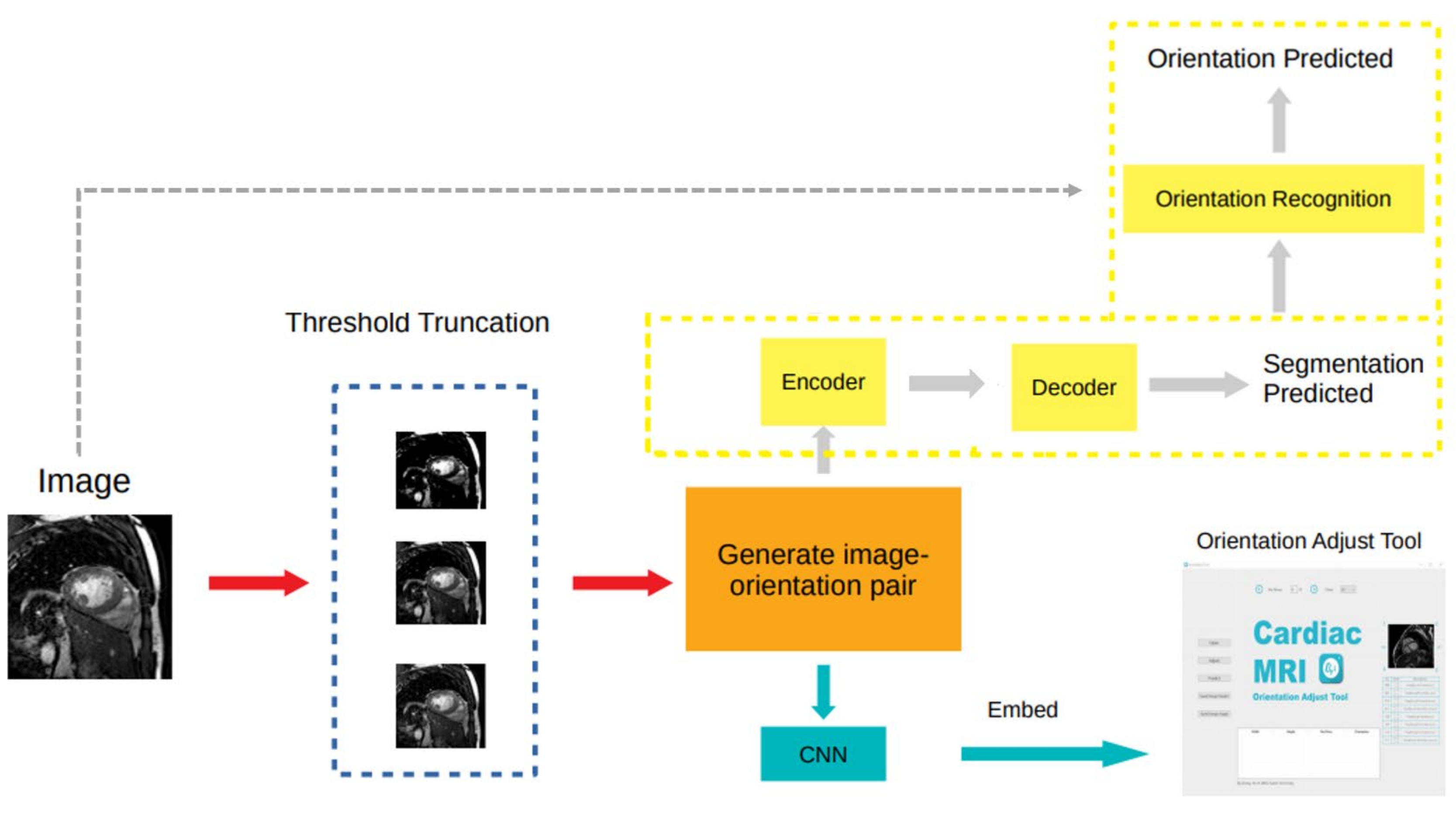}
\caption{The pipeline of the proposed CMR orientation recognition and standardization method. The image is first truncated at several gray value thresholds. Then the processed image is used to generate the image-orientation pair (see section2.1). Then, the multi-tasking network generates the orientation predicted and segmentation mask. When embed the orientation recognition network to the orientation adjust tool, the multi-tasking network is replaced with a simplified CNN. } \label{fig1}
\end{figure}

Given that many clinical applications rely on both an accurate segmentation and orientation recognition to extract specific anatomy or compute some functional indices, we therefore further propose a new multi-task learning framework that aims to solve the CMR orientation recognition and cardiac segmentation tasks at the same time. To enable the proposed model to be conveniently applied in medical image processing and clinical practice,  we develop a CMR Orientation Adjust Tool, resulting in a DNN model referred to as CMRadjustNet.
For simplicity, the CMR Orientation Adjust Tool is referred to as the CMRadjustNet Tool or CMRadjustNet in the remaining of the article.  CMRadjustNet is embedded with the proposed simplified orientation recognition network and support orientation recognition and standardization automatically. The tool has two versions, a graphical interface, and a command-line tool. The graphics interface version supports the display of MRI slices and orientation recognition, standardization, save functions.  We developed a user-friendly graphical interface to help users perform CMR image orientation standardization easily. The command-line tool supports batch orientation correction of all MRI files in a folder, which facilitates the processing of large amounts of MRI data.

This work is aimed at designing a DNN-based approach to achieve orientation recognition and standardization for multiple CMR modalities. Figure~\ref{fig1} presents the pipeline of our proposed method. The main contributions of this work are summarized as follows:
\begin{itemize}
    \item[(1)]We propose a scheme to standardize the CMR image orientation and categorize all the orientations for classification. 
    \item[(2)] We present a DNN-based orientation recognition method for CMR image and transfer it to other modalities.
    \item[(3)] We propose a multi-tasking network, where orientation recognition and image segmentation tasks are implemented simultaneously. 
    \item[(4)] We develop a CMR image orientation adjust tool (CMRadjustNet) embedded with a simplified orientation recognition network, which facilitates the CMR image orientation recognition and standardization in clinical and medical image processing practice. 
\end{itemize}


\section{Method}


In this section, we introduce our proposed method for orientation recognition and standardization. Our proposed framework is built on the categorization of CMR image orientations.  We propose a multi-tasking network and embed the simplified orientation recognition network into the CMR orientation adjust tool, i.e., CMRadjustNet.

\subsubsection{CMR Image Orientation Categorization}
Due to different data sources and scanning habits, the orientation of different cardiac magnetic resonance images may be different, and the orientation vector corresponding to the image itself may not correspond correctly. This may cause problems in tasks such as image segmentation or registration.
Taking a 2D image as an example, we set the orientation of an image as the initial image and set the four corners of the image to \framebox[1cm][c]{$\begin{array}{cc} 1 & 2 \\ 3 & 4 \end{array}$}, Then the orientation of the 2D MR image may have the following 8 variations, which is listed in Table ~\ref{tab1}. For each image label pair $X_t,\ Y_t$. One target orientation $O_t$ is randomly picked from the 9 orientation classes, correspondingly, we flip $X_t,\ Y_t$ towards the picked orientation. Then we obtain the image-label pair $X_t,\ Y_t$ and image-orientation pair $X_t,\ O_t$ . We denote the process of generating image-orientation pair as function $g$.

\begin{table}[thb]\centering
\caption{Orientation Categorization of 2D CMR Images. Here, sx, sy and sz respectively denote the size of image in X-axis, Y-axis and Z-axis.}\label{tab1} 
\begin{tabular}{|l| l c l |}
\hline
No. &	Operation &	\ \ \ Image \ \ \ &	Correspondence of coordinates\\
\hline
000 &	initial state & \framebox[0.8cm][c]{$\begin{array}{cc} 1 & 2 \\ 3 & 4 \end{array}$} & Target[x,y,z]=Source[x,y,z]\\ \hline
001 &	horizontal flip &\framebox[0.8cm][c]{$\begin{array}{cc} 2 & 1 \\ 4 & 3 \end{array}$}& Target[x,y,z]=Source[sx-x,y,z]\\\hline
010 &	vertical flip &\framebox[0.8cm][c]{$\begin{array}{cc} 3 & 4 \\ 1 & 2 \end{array}$}& Target[x,y,z]=Source[x,sy-y,z]\\\hline
011 &	Rotate $180^\circ$ clockwise &\framebox[0.8cm][c]{$\begin{array}{cc} 4 & 3 \\ 2 & 1 \end{array}$} & Target[x,y,z]=Source[sx-x,sy-y,z]\\\hline
100 &	
$\begin{array}{l}
\mathrm{Flip\ along\ the\ upper\ left\!-\!lower\ } \\\mathrm{right\ corner} \end{array}$
&\framebox[0.8cm][c]{$\begin{array}{cc} 1 & 3 \\ 2 & 4 \end{array}$}&Target[x,y,z]=Source[y,x,z]\\\hline
101 &	Rotate $90^\circ$ clockwise &\framebox[0.8cm][c]{$\begin{array}{cc} 3 & 1 \\ 4 & 2 \end{array}$}& Target[x,y,z]=Source[sx-y,x,z]\\\hline
110	& Rotate $270^\circ$ clockwise &\framebox[0.8cm][c]{$\begin{array}{cc} 2 & 4 \\ 1 & 3 \end{array}$}	& Target[x,y,z]=Source[y,sy-x,z]\\\hline
111	& 
$\begin{array}{l}
\mathrm{Flip\ along\ the\ bottom\ left\!-\!top\ } \\\mathrm{right\ corner} \end{array}$
&\framebox[0.8cm][c]{$\begin{array}{cc} 4 & 2 \\ 3 & 1 \end{array}$} & Target[x,y,z]=Source[sx-y,sy-x,z]\\
\hline
\end{tabular}
\end{table}

\subsubsection{Multi-tasking Network} 
Suppose given image-label pair $(X_t, Y_t)$, $X_t$ is then normalized. We denote the processed $X_t$ as $X'$. After generating image-orientation pair, $g(X')$ is taken as the input of multi-tasking network. Denote the encoder of proposed multi-tasking network as $encoder$, the decoder of proposed multi-tasking network as $decoder$, the model pipeline is formulated as below:

$$X_{feature}=encoder(g(X'))$$
$$\hat{Y}= decoder( X_{feature})$$
$$\hat{O}=F_{orientation}([g(X'), \hat{Y}]) .$$

Here, $\hat{Y}$ and $g(X')$ are concatenated and pass through orientation recognition sub-network $F_{orientation}$ to predict orientation classification. 
We start with the orientation recognition branch in the multi-tasking network, where we are interested in take segmentation masks predicted by segmentation network as attention map.
In the proposed multi-tasking framework, the orientation recognition sub-network consists of 3 convolution layer and a fully connected layers. 
The orientation predicted is denoted as $\hat{O}$. We use the standard categorical loss to calculate the loss between predicted orientation $\hat{O}$ and orientation label $O$,

$$L_{orientation} = \sum_{i=1}^{C}O_i log(\hat{O_i}) .$$
where,  $i$ denotes the orientation category. In our orientation classification setting, we set $C=8$.

Figure 2 shows the overall structure of the multi-tasking network. The backbone of the network is based on the Unet segmentation model. Since the network is originally designed to generate segmentation results, we keep the segmentation branch unchanged and add a new orientation recognition branch which take the segmentation mask as attention map. A weighted binary cross-entropy function is applied to calculate the multi-label segmentation loss between $\hat{Y_i}$ and $Y_i$.  $s$ is set as 4 when dealing with the CMR image segmentation task, which generates background, right ventricle, left ventricle, myocardium. The multi-label segmentation loss is formulated as below:

$$L_{segmentation}(\hat{Y},Y) = -\sum_{i}^{s}[Y_i log \hat{Y_i}+(1-Y_i)log(1-\hat{Y_i})]w_i . $$

By weighting Orientation recognition loss and segmentation loss, an integral loss function of the proposed loss function can be obtained,

$$L_{integral} = L_{segmentation} + L_{orientation}.$$

The proposed multitask network starts by learning the optimal segmentation network. Thus, $encoder$ and $decoder$ are optimized according to $L_{segmentation}$. Once this is complete, both $encoder$
and $decoder$ are fixed and the parameters of $F_{orientation}$ are re-initialized. Now, $F_{orientation}$ is trained according to $L_{orientation}$. Finally, we fine-tune the segmentation module and orientation recognition module simultaneously to obtain an optimized model on both segmentation task and orientation recognition task.

\subsubsection{CMR image orientation adjust tool} 

\begin{figure}[thb]
\includegraphics[width=\textwidth]{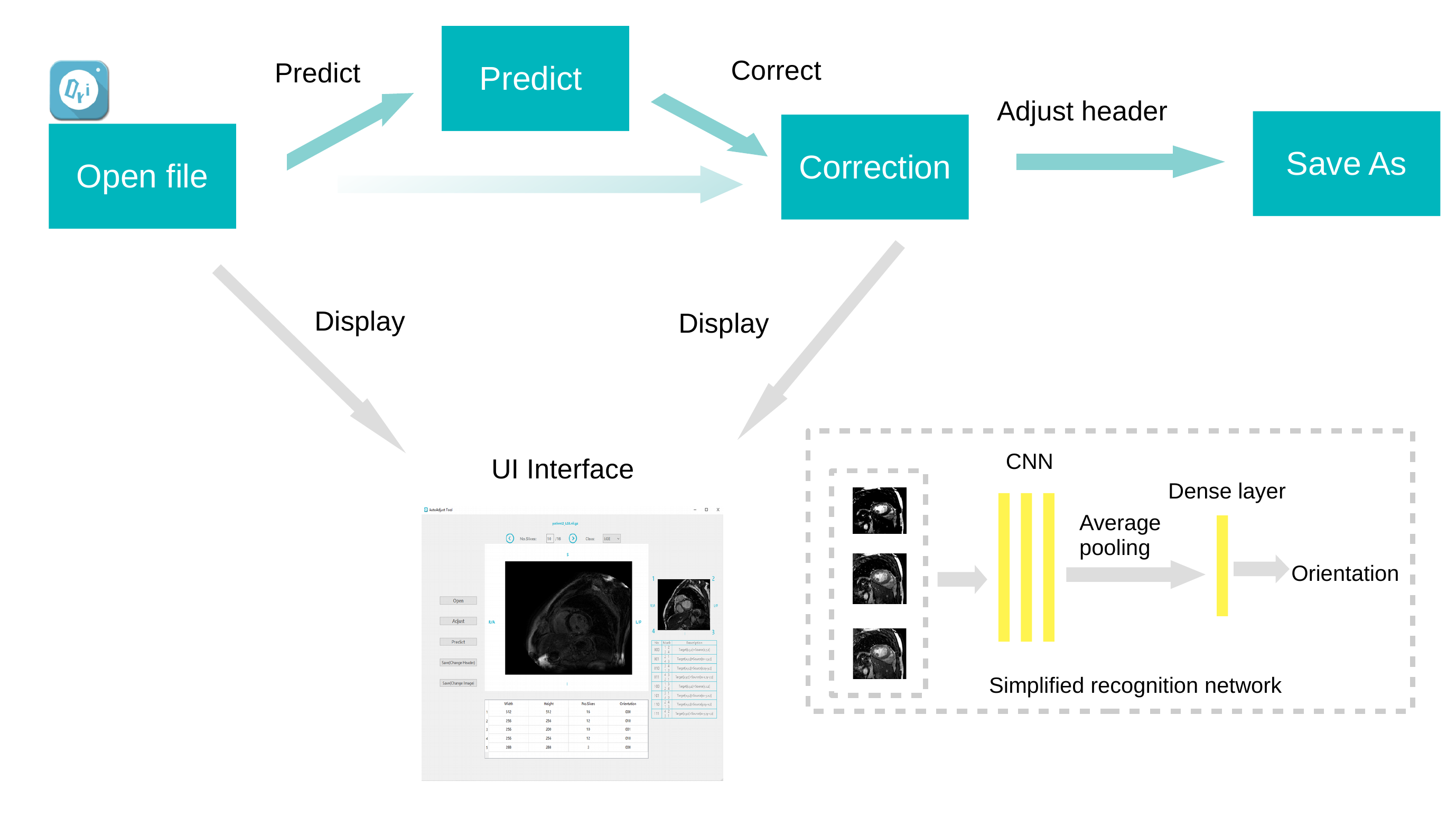}
\caption{The pipeline of the proposed CMR image orientation adjust tool.} \label{fig2}
\end{figure}

To visualize CMR images and perform image orientation recognition, standardization, save operations, we develop a DNN-based CMR image orientation adjust tool, which is embedded with an orientation recognition network.  To shorten the response time of the CMR Orientation Adjust Tool, we replace the multi-tasking network with a  simplified 3-layer CNN network. We also adopt a different preprocessing method. Suppose given image-label pair $(X_t, Y_t)$ , for each pair of $X_t$. We denote the maximum gray value as $G$. Three truncation operations are performed on $X_t$ at thresholds $60\%G, 80\%G, G$ to produce ${X_1}_t,{X_2}_t,{X_3}_t$ respectively. The truncation operation maps the pixel whose gray value higher than the threshold to the threshold gray value. Setting different thresholds enforces the characteristics of the image under different gray value window widths to avoid the influence of extreme gray values. The grayscale histogram equalization is also performed on ${X_1}_t,{X_2}_t,{X_3}_t$ to obtain ${X'_1}_t,{X'_2}_t,{X'_3}_t$. We found that the equalization preprocessing of the gray histogram can make the model converge more stably during training. We denote the concatenated 3-channel image $[{X'_1}_t, {X'_2}_t, {X'_3}_t]$ as $X’$. The orientation recognition CNN only retains the orientation recognition branch while keeping the image-orientation generation steps unchanged.  Figure ~\ref{fig2} presents the pipeline of the proposed CMR image orientation adjust tool.

When adapting the proposed orientation recognition network from a single modality to other modalities, we adopt a transfer learning method to obtain the transferred model. For example, we pre-train model on the balanced-Steady State Free Precession (bSSFP) cine dataset and then transfer model to late gadolinium enhancement (LGE) CMR or the T2-weighted CMR dataset.  We first fix the network parameters except for the encoder and retrain the connected layer on the new modality dataset. We go to the next fine-tune step until the model converges.  In the fine-tune training, we retrain the encoder and fully connected layer simultaneously on the new modality dataset to obtain an adapted model.

The graphical interface version is suitable for visualization and orientation standardization of a single CMR image. The complementary version of the graphical interface version is the command-line tool version, which supports batch orientation standardization operations of CMR images and provides a simple parameter setting method. By specifying a folder, one line of command is enough to identify the orientation of all MRI files in the folder and correct the files with the wrong orientation.

\section{Experiment}
\subsubsection{Experiment setup} We evaluate our proposed multi-tasking framework and orientation recognition network on the MyoPS dataset~\cite{ref_article7,ref_article6} and ACDC dataset~\cite{ref_article14}. The MyoPS dataset was similar to the previous challenge data, i.e., multi-sequence CMR segmentation (MS-CMRseg) challenge, of which both provide the three-sequence CMR (LGE, T2, and bSSFP) and three anatomy masks, including myocardium (Myo), left ventricle (LV), and right ventricle (RV). MyoPS further provides two pathology masks (myocardial infarct and edema) from the 45 patients. The ACDC dataset comprises of single modality short-axis cardiac cine-MRIs of 100 subjects from 5 groups - 20 normal controls and 20 each with 4 different cardiac abnormalities. Annotations are provided for LV, Myo and RV for both end-systole (ES) and end-diastole (ED) phases of each subject. We evaluate multi-task network on ACDC dataset. For the simplified orientation recognition network, we train model for single modality on the MyoPS dataset, then transfer the model to other modalities. For each sequence, we resample each slice of each 3d image and the corresponding labels to an in-plane resolution of $1.367\times1.367$ mm.  Image slices are cropped or padded to $212\times 212$ for multi-task network and resized to $100\times 100$  For the simplified orientation recognition network. We divide all slices into three sub-sets, i.e., the training set, validation set, and test set, at the ratio of $80\%$, $10\%$, and $10\%$.  

\subsubsection{For multi-tasking network} 
The performance of the orientation recognition module was evaluated by using the accuracy between the predicted orientation and the target orientation. Dice Score was used to measure the accuracy of segmentation. Dice score is an ensemble similarity measurement function, which is usually used to calculate the similarity of two samples. For the predicted segmentation result $\hat{Y}$ and ground truth $Y$, the dice score is formulated as follows,
$$s = \frac{2|\hat{Y} \cap Y|}{|\hat{Y}|+|Y|} .$$

\begin{table}[thb]
\caption{Segmentation Dice Score and orientation recognition accuracy for multi-tasking network. Note: reported dice score are the average(standard deviation in parenthesis).}\label{tab3} 
\begin{center}
\begin{tabular}{|l| c c c  | c |}
\hline
\multirow{2}{*}{Tasks} 
& \multicolumn{3}{c}{Segmentation (Dice)} &  Orientation recognition\\ 
&	LV & MYO &RV & Accuracy\\
\hline
Results & 0.920(0.11)& 0.853(0.05) & 0.757(0.14) & 0.987 \\
\hline
\end{tabular}
\end{center}
\end{table}


Table~\ref{tab3} presents quantitative results of our proposed multi-task network. It can be observed that the proposed method achieves a good segmentation result, with the average dice score of 0.843. The accuracy of orientation recognition  reaches 0.986. The quantitative results prove that the proposed multi-task model can effectively deal with the segmentation  and orientation recognition tasks simultaneously.

\subsubsection{For simplified orientation recognition network} 
In each training iteration, a batch of the three-channel images $X’$ is fed into the simplified orientation recognition network (see Figure 2). Then, the network outputs the predicted orientation network, which is denoted as a $1\times 3$ vector. The predicted orientation is then fed into the standardization module of the CMR image orientation adjust tool. The inverse operation of the predicted orientation error is performed, which warps the MR image to the correct orientation. Then, we perform the same inverse operation on the Orientation vector in the MR file header to obtain the MR file with correct orientation. 

\begin{table}[thb]
\caption{Orientation recognition accuracy of 2D MS-CMR.}\label{tab2}
\begin{center}
\begin{tabular}{lll}
\hline
Modality &	Accuracy &	Description\\
\hline
bSSFP & 0.990 & pre-train \\
LGE & 0.852 & transfer learning \\
T2 &  0.980 & transfer learning \\
\hline
\end{tabular}
\end{center}
\end{table}

Table~\ref{tab2} shows the average accuracy on the data set. The description indicates whether the model was trained on this modality or was transferred from other modalities. The high accuracy results provide us with the necessary conditions for the development of the CMR image orientation adjust tool.

\section{Conclusion}
We have proposed a multi-tasking framework for multi-sequence MRI images that deal with segmentation and orientation recognition tasks simultaneously. Also, we have developed the CMR Orientation Adjust tool (CMRadjustNet), which is embedded with a simplified orientation recognition network. The experiment demonstrates that the embedded orientation recognition network is capable of recognizing the orientation classification from multi-sequence CMR images. Our future research aims to expand the categorization of the CMR image orientation, and study orientation standardization on 3D MRI images.

%
%
%
\bibliographystyle{splncs04}
\bibliography{paper27.bib}

\end{document}